%% file: paper_la.tex
\documentclass[twocolumn,showpacs,aps,prl,superscriptaddress,floatfix]{revtex4}
\usepackage{epsfig}
\usepackage{graphicx}
\usepackage{dcolumn}
\usepackage{amsmath}
\usepackage{multirow}

\RequirePackage{xspace}
\input pubboard/babarsym

\newcommand{\BABARPubYear}     {09}
\newcommand{\BABARPubNumber}   {010}
\newcommand{\SLACPubNumber} {13674}

\def\kszkpbr{4.45}
\def\kszkpbrstat{0.10}
\def\kszkpbrsyst{0.17}

\def\kszkszbr{4.66}
\def\kszkszbrstat{0.37}
\def\kszkszbrsyst{0.35}

\def\kspkpzbr{4.38}
\def\kspkpzbrstat{0.19}
\def\kspkpzbrsyst{0.26}

\def\kspkspbr{4.13}
\def\kspkspbrstat{0.18}
\def\kspkspbrsyst{0.16}

\def\kszcombr{4.47}
\def\kszcomstat{0.10}
\def\kszcomsyst{0.16}

\def\kspcombr{4.22}
\def\kspcomstat{0.14}
\def\kspcomsyst{0.16}

\def\isoval{0.066}
\def\isostat{0.021}
\def\isosyst{0.022}

\def\isolow{0.017}
\def\isohi{0.116}

\def\acpkp{0.018}
\def\acpkpstat{0.028}
\def\acpkpsys{0.007}

\def\acpval{-0.003}
\def\acpstat{0.017}
\def\acpsyst{0.007}

\def\acplolim{-0.033}
\def\acphilim{0.028}

\def\ifb{\;\mbox{fb}^{-1}}
\def\lumitotal{347\ifb}

\def\numBB{383 million}

\def\GeV{\;\mbox{GeV}}

\def\GeVcc{\;\mbox{GeV}/c^2}

\def\de   {\Delta E}
\def\dz   {\Delta z}
\def\mes  {m_{\mbox{\scriptsize ES}} }
\def\bkg  {B \to K^{*}\gamma}

\def\bkgneut  {B^0 \to K^{*0}\gamma}
\def\bkpg  {B^+ \to K^{*+}\gamma}

\def\bsg    {b\to s\gamma}

\def\kszkp  {K^{*0}\! \rightarrow \! K^+\pi^-}
\def\kspkpz {K^{*+}\! \rightarrow \!  K^+\pi^0}
\def\kspksp {K^{*+} \! \rightarrow \! K_S\pi^+}
\def\kszksz {K^{*0} \! \rightarrow \! K_S\pi^0}

\def\incbsg  {B\to X_s\gamma}

\def\acp {\ensuremath{\mathcal{A}}}
\def\aexcl {\ensuremath {  \acplolim <\acp(\bkg) < \acphilim }}
\def\isoexcl{\ensuremath {\isolow < \Delta_{0-} < \isohi }}

\def\isoas{\Delta_{0-}}

\def\B       {\ensuremath{B}\xspace}
\def\Bbar    {\kern 0.18em\overline{\kern -0.18em B}{}\xspace}

\def\BB      {\ensuremath{B\Bbar}\xspace} 

\def\bkg    {\ensuremath {\B \to \Kstar \gamma}}

\def\bkpg    {\ensuremath {\Bp \to \Kstarp \gamma}}

\def\Kz    {\ensuremath{K^{0}}\xspace}

\def\CP                {\ensuremath{C\!P}\xspace}

\begin{document}

\preprint{\babar-PUB-\BABARPubYear/\BABARPubNumber}
\preprint{SLAC-PUB-\SLACPubNumber}

{
\pagestyle{empty}
\begin{flushleft}
\babar-PUB-\BABARPubYear/\BABARPubNumber\\
SLAC-PUB-\SLACPubNumber\\
\end{flushleft}
}

\title{\large\bf\boldmath
Measurement of Branching Fractions and \CP and Isospin Asymmetries in $B\to K^*(892)\gamma$
Decays
}

\collaboration{The \babar\ Collaboration}
\noaffiliation

\date{\today}

\begin{abstract}
We present an analysis of the decays $B^0 \to K^{*0}(892)\gamma$ and $B^+ \to K^{*+}(892)\gamma$ 
using a sample of about
\hbox{\numBB~\BB\  }events 
collected with the \babar\ detector at the PEP-II asymmetric energy $B$ factory.  
We measure the branching fractions 
\hbox{${\cal B}(\bkgneut) =  (\kszcombr \pm \kszcomstat \pm \kszcomsyst)\times 10^{-5}$  }
and 
\hbox{${\cal B}(\bkpg) =  (\kspcombr \pm \kspcomstat \pm \kspcomsyst) \times 10^{-5}$.}  
We constrain the direct \CP asymmetry to be \aexcl~and the isospin asymmetry to be 
\hbox{\isoexcl}, where the limits are determined by the 90\% confidence interval 
and include both the statistical and systematic uncertainties.
\end{abstract}

\input pubboard/authors_mar2009

\maketitle

In the Standard Model (SM), the decays \mbox{\bkg}~\cite{Kstref} proceed dominantly 
through one-loop \mbox{$\bsg$} electromagnetic penguin transitions. 
Some extensions of the SM predict new high-mass particles that can exist in 
the loop and alter the branching fractions from their SM predictions.
Previous measurements of the branching fractions~\cite{Coan:1999kg,  Dasu:2004kg, 
Nakao:2004kg} are in agreement with and more precise 
than SM predictions~\cite{Ali:2001ez, Bosch:2001gv, Beneke:2001at, Matsumori:2005ds, Ali:2008kg},
which suffer from large hadronic uncertainties.

The time-integrated \CP (\acp) and isospin ($\Delta_{0-}$) asymmetries have 
smaller theoretical uncertainties~\cite{CC}, and therefore provide more 
stringent tests of the SM.  They are defined by:

\begin{equation}
\acp = 
{
{\Gamma(\overline{B} \to \overline{K}^*\gamma) - \Gamma(\bkg)} \over
{\Gamma(\overline{B} \to \overline{K}^*\gamma) + \Gamma(\bkg)}
},
\label{eq:acp}
\end{equation}
\begin{equation}
\Delta_{0-} =
{
{\Gamma(\overline{B}^0 \to \overline{K}^{*0}\gamma) - \Gamma(B^{-}\to K^{*-}\gamma)} \over
{\Gamma(\overline{B}^0 \to \overline{K}^{*0}\gamma) + \Gamma(B^{-}\to K^{*-}\gamma)}
},
\label{eq:isospin}
\end{equation}

\noindent
where the symbol $\Gamma$ denotes the partial width.  The SM predictions for $\acp$ are on the order of 1\% ~\cite{Greub:1994},
while those for $\Delta_{0-}$ range from 2 to 10\%~\cite{Matsumori:2005ds, Kagan:2001zk}.  
However, new physics could alter these parameters 
significantly~\cite{Kagan:2001zk, Ahmady:2006gh, Dariescu:2007}, and thus
precise measurements can constrain those models. 
In this letter, we report measurements of
${\cal B}(\bkgneut)$, ${\cal B}(\bkpg)$, $\isoas$, and $\acp$.
We use a data sample containing about \numBB~\BB\ events, corresponding to an 
integrated luminosity of $\lumitotal$, recorded at a center-of-mass (CM) energy corresponding to the
$\Upsilon(4S)$ mass.  
The data was taken with the \babar\ detector~\cite{BabarDet} at the PEP-II asymmetric $e^+e^-$ collider.  
We also make use of events simulated using Monte Carlo (MC) methods and
a GEANT4~\cite{geant} detector simulation.
These results supercede the previous \babar\ measurements~\cite{Dasu:2004kg}.

$\bkg$ decays are reconstructed in the following $K^*$ modes:  
$\kszkp$, $\kszksz$, $\kspkpz$, and $\kspksp$.
For each signal decay mode, the selection requirements described below 
have been optimized for the maximum statistical sensitivity of $S/\sqrt{S+B}$, 
where $S$ and $B$ are the rates for signal and background, respectively, where the assumed 
signal branching fraction is $4.0\times 10^{-5}$~\cite{Dasu:2004kg}.
The dominant source of background is continuum events 
($\ep\en \to q\bar{q}(\gamma)$, with $q=u,d,s,c$) 
that contain a high-energy photon from a $\piz$ or $\eta$ decay or
from an initial-state radiation (ISR) process.
Backgrounds coming from \BB events are mostly from 
higher-multiplicity $\b\rightarrow\s\gamma$ decays, 
where one or more particles have not been reconstructed, and from 
decays of one $\bkg$ mode that enter the signal selection 
of another mode by  mis-reconstructing the $K^*$ meson.

Photon candidates are identified as localized energy deposits in the 
calorimeter (EMC) that are not associated with any charged track.
The signal photon candidate is required to have a CM energy between 
1.5 and 3.5 \gev, to be well-isolated and to have a shower shape
consistent with an individual photon~\cite{BabarksgOld}.  
In order to veto photons from $\piz$ and $\eta$ decays,
we form photon pairs composed of the signal photon candidate and all 
other photon candidates in the event.  
We then reject signal photon candidates consistent  
with coming from a $\piz$ or $\eta$ decay 
based on a likelihood ratio that uses the energy of
the partner photon, and the invariant mass of the pair.

Charged particles, except those used to form $K_S$ candidates, are selected 
from well-reconstructed tracks that have at
least 12 hits in the Drift Chamber (DCH), and are required to be consistent with coming 
from the \ep\en interaction region.  
They are identified as $K$ or $\pi$ mesons by the Cherenkov angle measured
in the Cherenkov photon detector (DIRC) as well as by energy loss of the track 
(\dedx) in the silicon vertex tracker (SVT) and DCH.  
The $K_S$ candidates are reconstructed from two oppositely charged tracks 
that come from a common vertex.  
In the $\kszksz$ ($\kspksp$) mode, we require the invariant mass of the pair to be 
$0.49 < m_{\pi^{+}\pi^{-}} < 0.52\GeVcc$ ($0.48 < m_{\pi^{+}\pi^{-}} < 0.52\GeVcc$) 
and the reconstructed decay length of the $K_S$ to be at least 9.3(10) times its uncertainty.  

We form $\piz$ candidates by combining two photons (excluding the signal photon candidate) in the event, each of which has an energy greater than 30 \mev in the laboratory frame.  
We require the invariant mass of the pair to be 
$0.112 < m_{\gamma\gamma} < 0.15\GeVcc$ and 
$0.114 < m_{\gamma\gamma} < 0.15\GeVcc$ for the $\kszksz$ and $\kspkpz$ modes respectively.    
In order to refine the $\piz$ three-momentum vector, we perform a mass-constrained fit 
of the two photons.

We combine the reconstructed $K$ and $\pi$ mesons to form $K^*$ candidates.  
We require the invariant mass of the pair to satisfy 
$0.78 < m_{K^{+}\pi^{-}} < 1.1\GeVcc$, 
$0.82 < m_{K_{S}\pi^{0}} < 1.0\GeVcc$, 
$0.79 < m_{K^{+}\pi^{0}} < 1.0\GeVcc$, 
and $0.79 < m_{K_{S}\pi^{+}} < 1.0\GeVcc$. 
The charged track pairs of the $\kszkp$ mode are required to originate 
from a common vertex.   

\begin{figure*}[htpb]

\includegraphics[width=0.245\linewidth,clip=true]{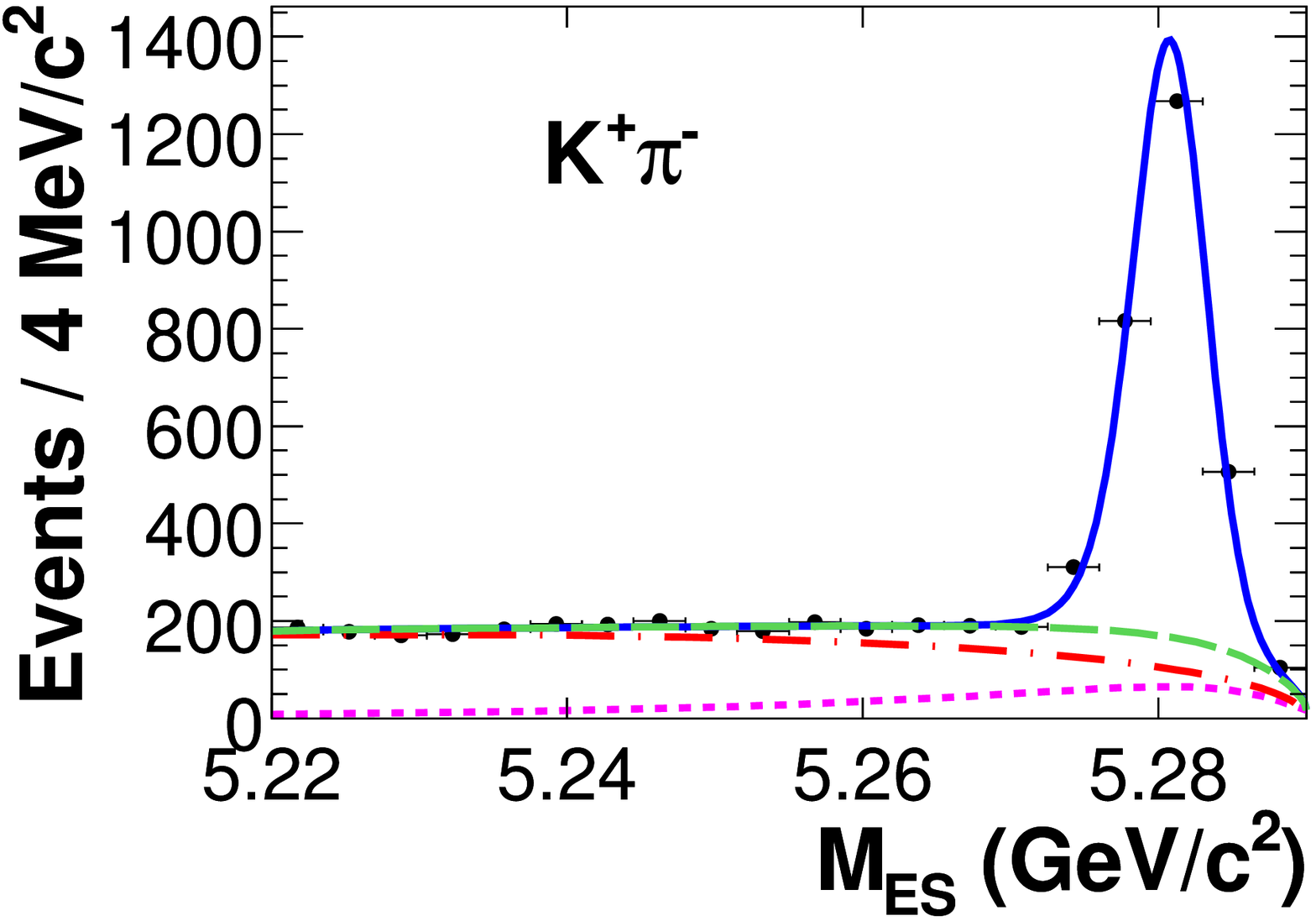}%
\includegraphics[width=0.245\linewidth,clip=true]{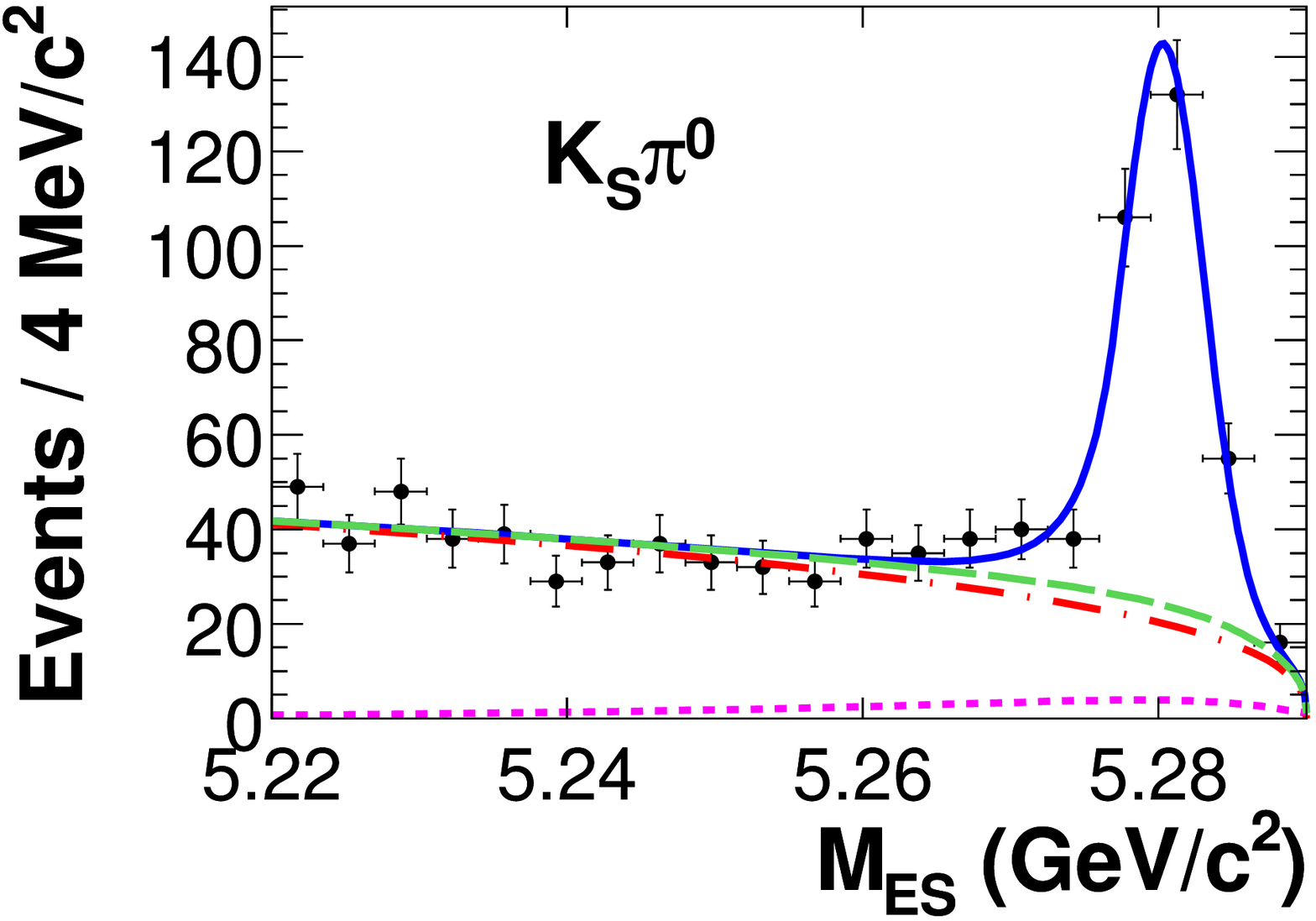}
\includegraphics[width=0.245\linewidth,clip=true]{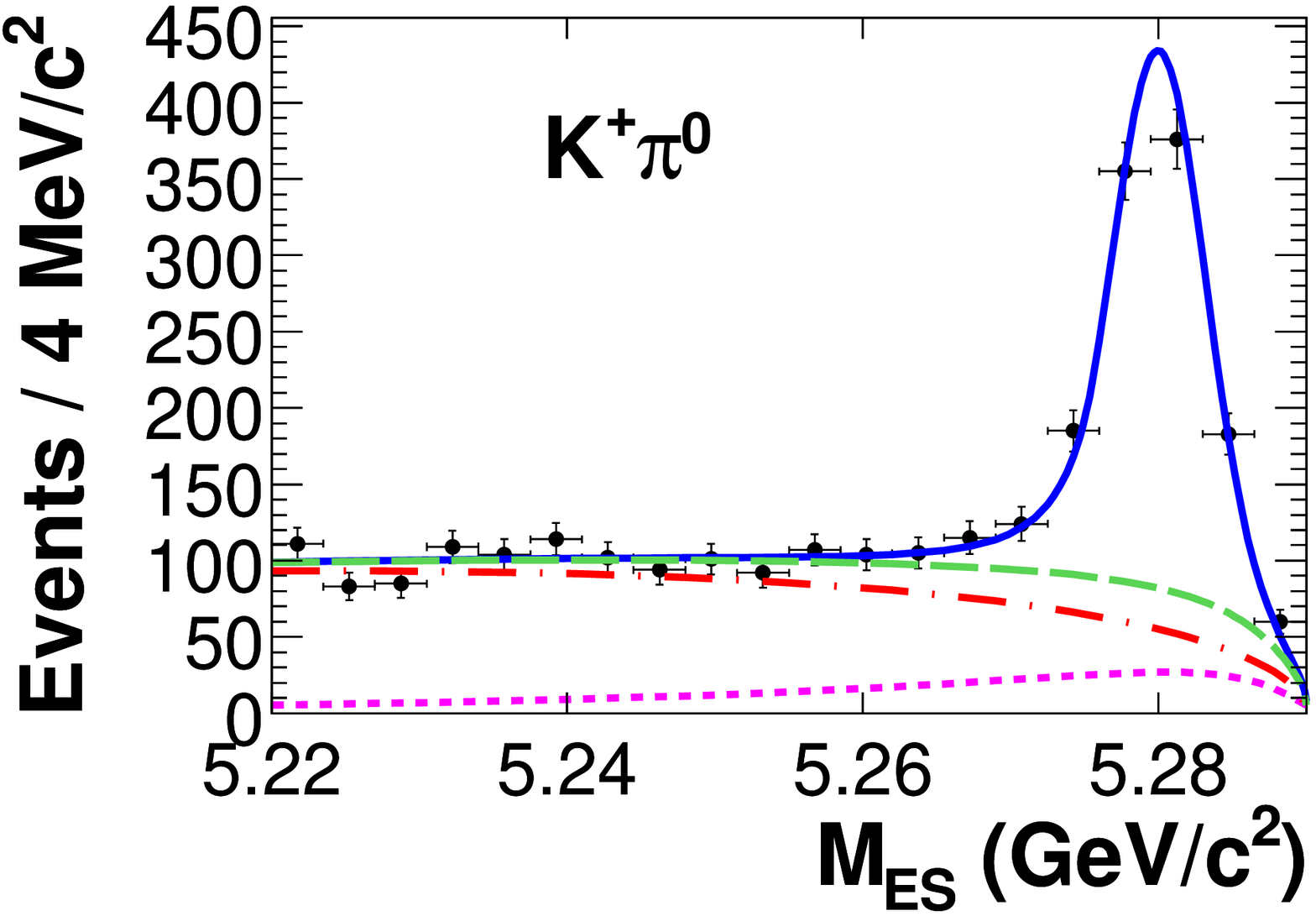}%
\includegraphics[width=0.245\linewidth,clip=true]{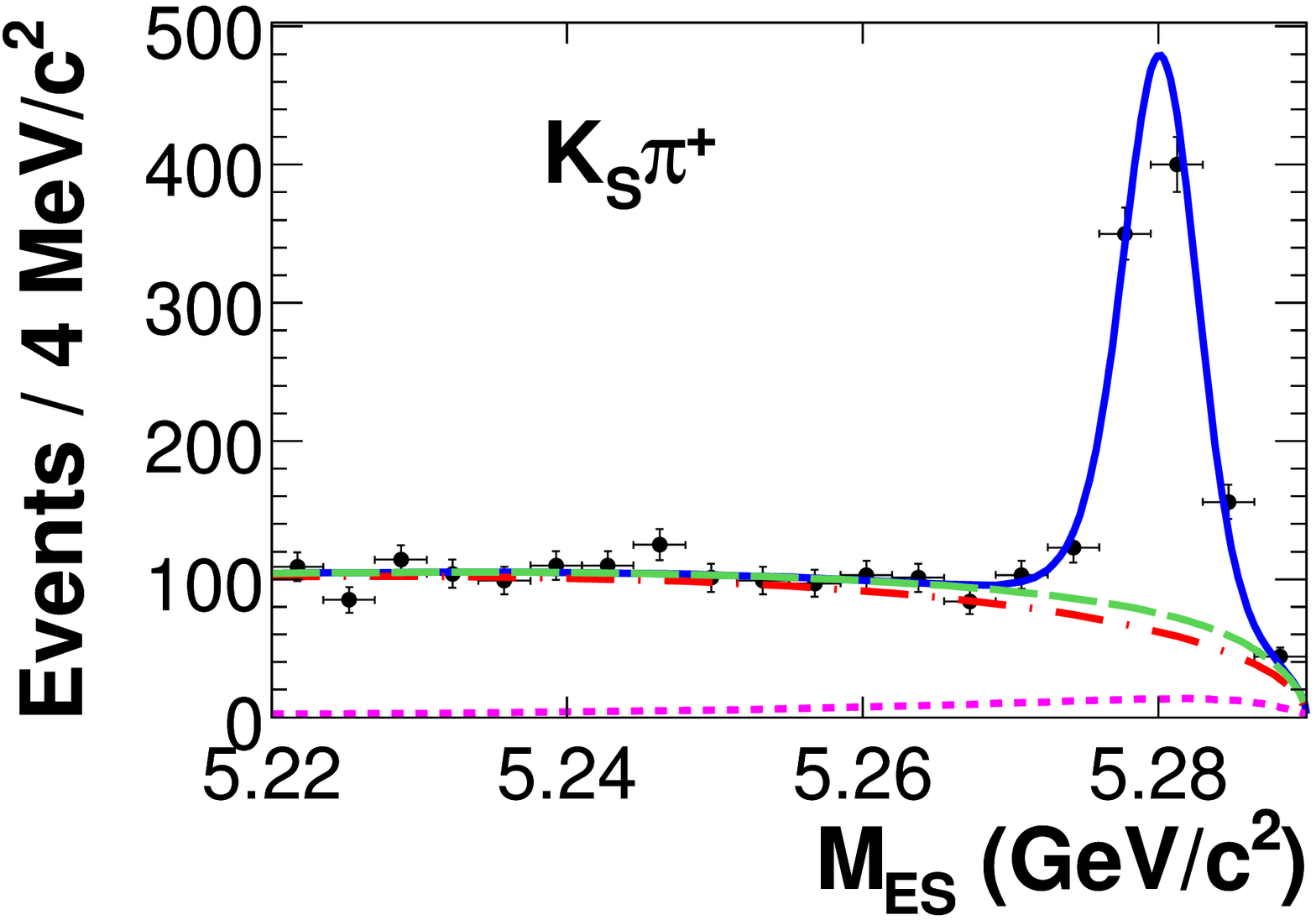}

\includegraphics[width=0.245\linewidth,clip=true]{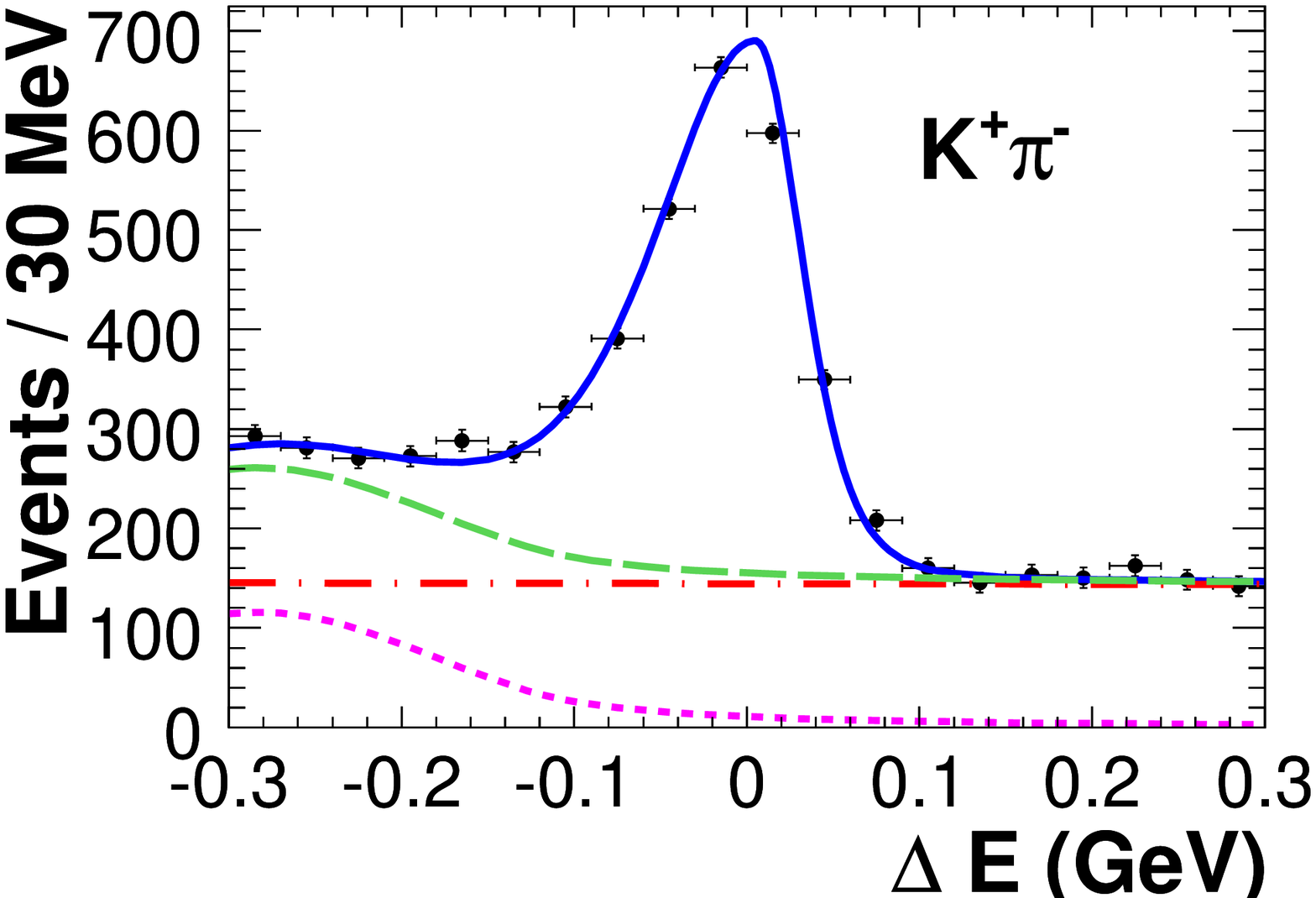}%
\includegraphics[width=0.245\linewidth,clip=true]{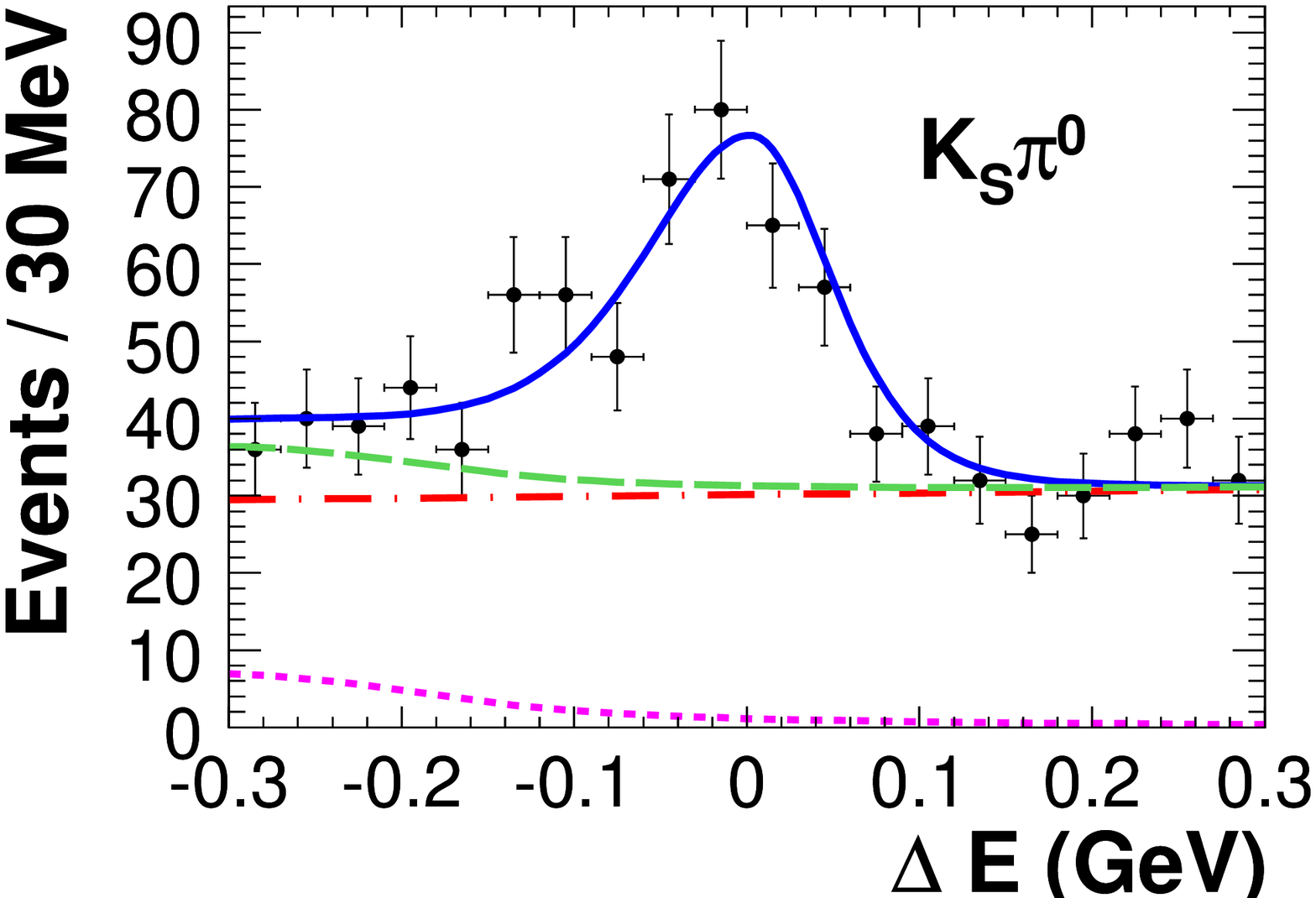}
\includegraphics[width=0.245\linewidth,clip=true]{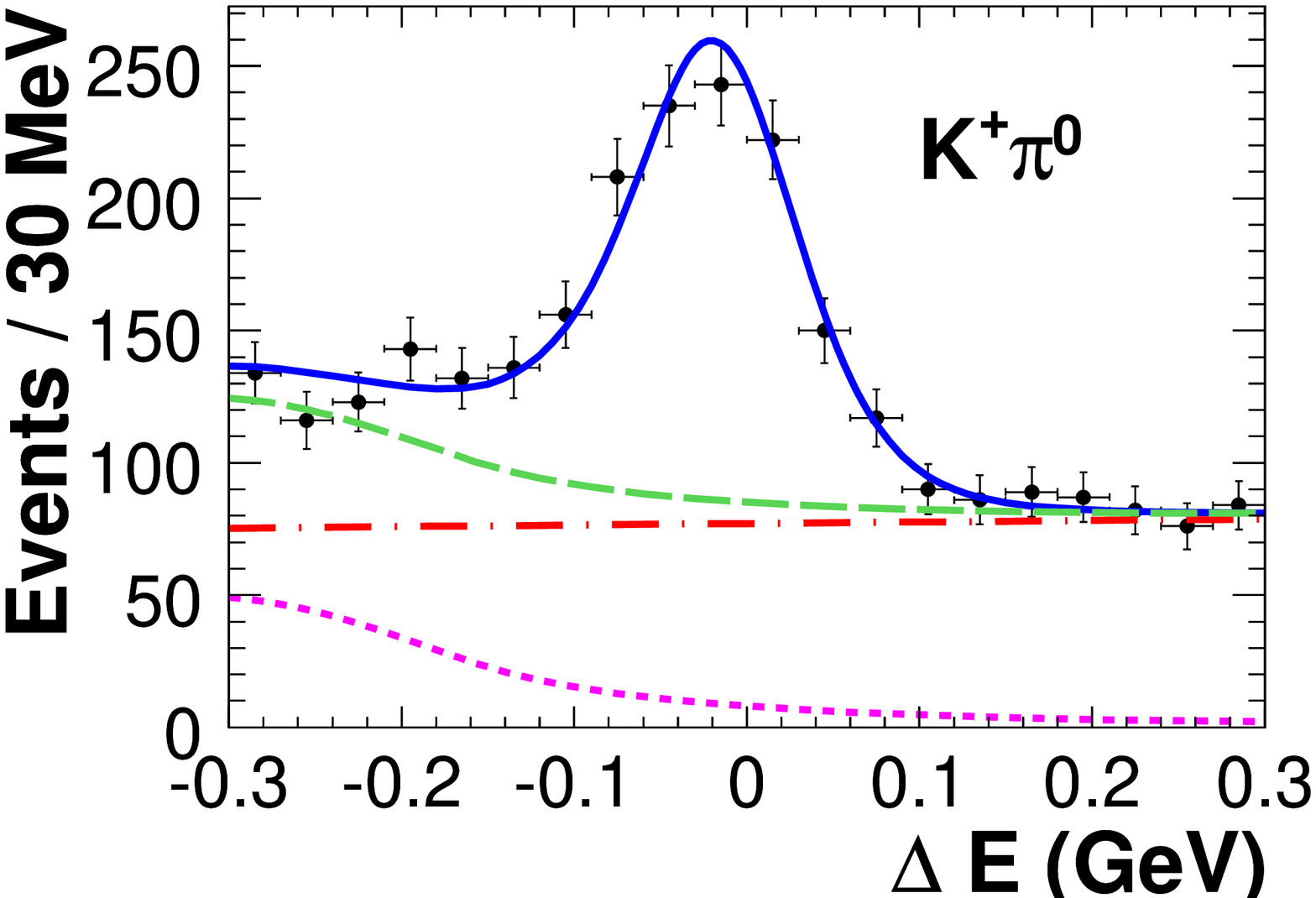}%
\includegraphics[width=0.245\linewidth,clip=true]{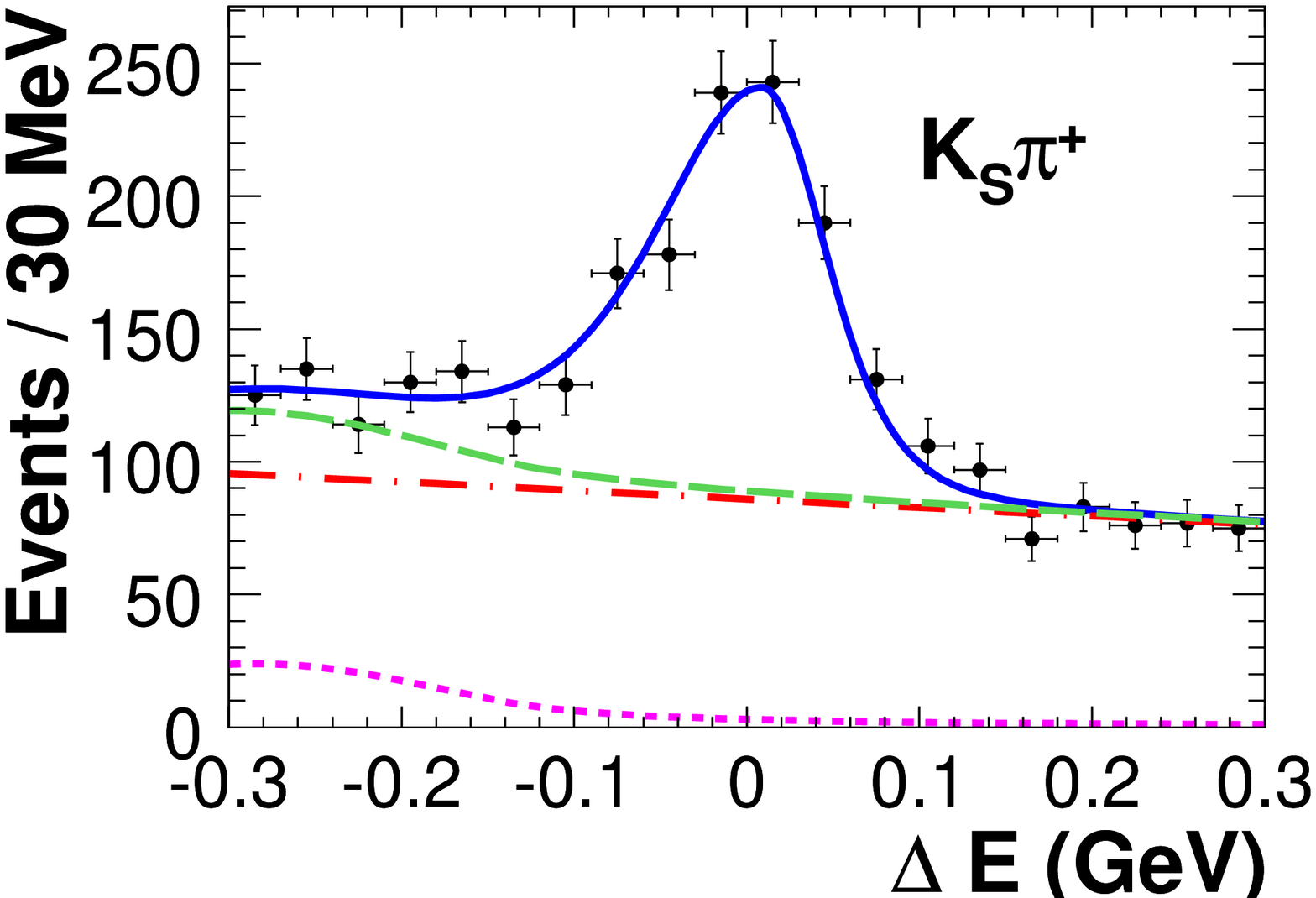}

\caption{$\mes$ and $\de$ projections of the fits. The points are data, the solid line is the fit result, the dotted line is the $B\overline{B}$ background, and the dash-dotted line is the continuum background.  The dashed line gives the total ($B\overline{B}$ and continuum) contribution to the background.  
}\label{fig:kstarfit}
\end{figure*}

The $K^*$ and high-energy photon candidates are combined to form $B$ candidates. 
We define in the CM frame (the asterisk denotes a CM quantity) 
$\de \equiv E^*_{B}-E_{\rm beam}^*$, where $E^*_B$ is the energy of the 
$B$ meson candidate and $E_{\rm beam}^*$ is the beam energy.  
The beam-energy-substituted mass is defined as 
$\mes \equiv \sqrt{ E^{*2}_{\rm beam}-{\mathrm{{p}}}_{B}^{\;*2}}$, 
where ${\mathrm{{p}}}_B^{\;*}$ is the momentum of the $B$ candidate.  
In addition, we consider the helicity angle $\theta_{H}$ of the $K^*$, 
defined as the angle between the momenta one of the daughters of the 
$K^*$ meson and the $B$ candidate in the $K^*$ rest frame.  The distribution of  
$\cos\theta_{H}$ is $\sin^2\theta$ for signal events.       
Signal events have $\de$ close to zero 
with a gaussian resolution of approximately $50\mev$, and an $\mes$ distribution 
centered at the mass of the $B$ meson with a gaussian resolution of 
approximately $3~\mevcc$.  
We only consider candidates in the ranges $-0.3 < \de <0.3 \GeV$, 
$\mes >  5.22\GeVcc$, and $|\cos\theta_{H}|< 0.75$.  
To eliminate badly reconstructed events, we apply
a loose selection criterion to the vertex separation (and its uncertainty) along the beam axis 
between the $B$ meson candidate and the rest of the event (ROE).  
The ROE is defined as all charged tracks and neutral energy deposits in 
the calorimeter that are not used to reconstruct the $B$ candidate.

\begin{table*}[htpb]
\caption{The signal reconstruction efficiency $\epsilon$, the fitted signal yield $N_S$,  
branching fraction, ${\cal B}$, and \CP asymmetry, $\acp$, for each decay mode.  
Errors are statistical and systematic, with the exception of $\epsilon$ and $N_S$, 
which have only systematic and statistical errors, respectively. }

        \begin{center}

        \begin{tabular*}{\linewidth}{
@{\extracolsep{\fill}}l
@{\extracolsep{\fill}}c
@{\extracolsep{\fill}}c
@{\extracolsep{\fill}}c
@{\extracolsep{\fill}}c
@{\extracolsep{\fill}}c
@{\extracolsep{\fill}}c
}\hline\hline

    Mode&
    {$\epsilon$(\%)}&
    {$N_S$}&
    {${\cal B}(\times\mbox{10}^{-\mbox{5}}) $}&
    {Combined ${\cal B} (\times\mbox{10}^{-\mbox{5}}) $}&
    {${\acp}$}&
    {Combined ${\acp}$}\\\hline

    $K^+\pi^-$&
    21.8$\pm$0.8&
    2400.0$\pm$55.4&
    $\kszkpbr \pm \kszkpbrstat \pm \kszkpbrsyst$&
    \multirow{2}*{$\Bigr{\}}\kszcombr \pm \kszcomstat \pm \kszcomsyst$}&
    $-0.016 \pm 0.022 \pm 0.007$&
    \multirow{4}*{$\Biggr{\}}\acpval \pm \acpstat \pm \acpsyst$}\\

    $K_s\pi^0$&
    13.0$\pm$0.9&
    \ 256.0$\pm$20.6&
    $\kszkszbr \pm \kszkszbrstat \pm \kszkszbrsyst$&&&\\

    $K^+\pi^0$&
    15.3$\pm$0.8&
    \ 872.7$\pm$37.6&
    $\kspkpzbr \pm \kspkpzbrstat \pm \kspkpzbrsyst$&
    \multirow{2}*{$\Bigr{\}}\kspcombr \pm \kspcomstat \pm \kspcomsyst$}&
    $+0.040 \pm 0.039 \pm 0.007$&\\

    $K_s\pi^+$&
    20.1$\pm$0.7&
    \ 759.1$\pm$33.8&
    $\kspkspbr \pm \kspkspbrstat \pm \kspkspbrsyst$&
    &
    $-0.006 \pm 0.041 \pm 0.007$&\\

    \hline\hline

    \end{tabular*}
    \end{center}

\label{table:results}
\end{table*}

In order to reject continuum background, we combine 13
variables into a neural network (NN).  
One class of these variables exploits the topological differences 
between isotropically distributed signal events and jet-like continuum events by 
considering correlations between the $B$ meson candidate and the ROE.
The other class exploits the fact that $B$ meson decays tend 
to not conserve flavor, while continuum events tend to be flavor-conserving.     
The discriminating variables are described in 
Ref.~\cite{rhogamma}.  
Each signal mode has a separately trained neural network, whose output peaks at a 
value of one for signal-like events and zero for background-like events.  A selection is made upon the output.

After applying all the selection criteria, there are, on average, $\sim1.1$ $B^0/B^+$ candidates 
per event in simulated signal events.  
In events with multiple candidates, we select the candidate with the reconstructed
$K^*$ mass closest to the nominal $K^*$ mass~\cite{PDG:2008}.

We perform an unbinned extended maximum likelihood fit to extract 
the signal yield, constructing a separate fit for each mode.  Since the correlations among the three observables $(\mes,\de,\cos\theta_{H})_j$ are small, we use uncorrelated probability distribution functions (PDFs) each representing the observables to construct the likelihood function.  The likelihood function is:

$$ 
{\cal L}=\exp{\left(-\sum_{i = 1}^{M} n_{i}\right)}\cdot
\left(\prod_{j = 1}^{N} \left[\sum_{i=1}^M n_i{\cal P}_i(\vec{x}_j;\
\vec{\alpha}_i)\right]\right)
 $$

\noindent
where $N$ is the number of events, $M = 3$ is the number of hypotheses (signal, continuum, and $\BB$), and $n_{i}$ is the yield of a particular hypothesis. ${\cal P}_i$ is the product of one-dimensional PDFS over the three dimensions $\vec{x}$, and $\vec{\alpha}$ represents the fit parameters.  
All types of $\BB$ background are included in the $\BB$ component, 
which is suppressed by the use of $\cos\theta_{H}$.
The signal $\mes$ distribution for the $\kszksz$ and $\kspkpz$ modes is
described by a Crystal Ball function~\cite{CrysBall}, which has two tail parameters that are fixed to values obtained from MC.
For the $\kszkp$ and $\kspksp$ modes, the signal $\mes$ distribution is parameterized as a piece-wise function 
$f(x) = \exp \left( -(x-\mu)^2/ (\sigma^2_{L,R} + \alpha_{L,R} (x-\mu)^2)
\right)$, defined to the left (L) and right (R) of $\mu$, which is the peak position of the distribution.  Here, $\sigma_{L,R}$ and $\alpha_{L,R}$ are the widths and measures of the tails, respectively, to the left
and right of the peak.
We constrain $\sigma_{L} = \sigma_{R}$, which is floated, and fix $\alpha_{L,R}$ to values obtained from MC.  
This same function also describes the signal $\de$ distribution 
for each mode, but with different values for the parameters.  
In addition, we allow $\sigma_{L}$ and $\sigma_{R}$ to float independently.
The $\cos\theta_{H}$ distribution for the signal component is
modeled by a $2^{\mathrm{nd}}$ order polynomial, with all of its parameters floating
in the fit.
For the continuum hypothesis, the $\mes$ PDF is parameterized
by an ARGUS function~\cite{argus}, with its shape parameter 
floating in the fit.  
The continuum $\de$ and $\cos\theta_{H}$ shapes are modeled 
by a first- or second-order polynomial with its parameters
floating in the fit.
Various functional forms are used to describe the $\BB$ background,
all parameters of which are taken from MC simulation and held fixed.  All of the component yields are floating.

Figure~\ref{fig:kstarfit} and Table~\ref{table:results} show
the results of the likelihood fit to data.  The branching fractions have been obtained using 
$\BR(\Upsilon (4S)\to\BzBzb) = 0.484 \pm 0.006,~\BR(\Upsilon (4S)\to\BpBm) = 
0.516 \pm 0.006$~\cite{PDG:2008}.    
Also shown are the combined branching fractions, which have been calculated 
taking into account correlated systematic errors.

The \CP asymmetry $\acp$ is measured in three modes:
$\kszkp$, $\kspkpz$, and $\kspksp$.  
In each of these modes, the final state of the signal $B$ meson is determined by its final state daughters.  The fit is accomplished by performing a simultaneous fit to the two flavor 
sub-samples ($K^*$ and $\overline{K}^*$) in each mode.
All shape parameters are assumed to be flavor independent and the $\acp$
of each component is floated in the fit. Table~\ref{table:results} gives the individual and combined $\acp$ results. 

Table~\ref{table:sys} lists the sources of systematic uncertainty 
for the branching fractions for all four modes.   
The ``Fit Model'' systematic incorporates uncertainties 
due to imperfect knowledge of the 
normalization and shape of the inclusive $\incbsg$ spectra, 
and the choice of fixed parameters. 
The ``Signal PDF bias'' systematic uncertainty characterizes any bias resulting 
from correlations among the three observables, or incorrect
modeling of the signal PDFs.
The remaining sources of error on the signal efficiency are studied using control
samples in the data. 
From all of these studies, we derive signal efficiency correction factors and associated uncertainties.
The total corrections are 0.953, 0.897, 0.919, and 0.936 for the 
$\kszkp$, $\kszksz$, $\kspkpz$, and $\kspksp$ modes, respectively. 
The systematic error on \acp\ comes largely from the uncertainty in the
charge asymmetry of the hadronic interaction of the final state mesons
with the detector material. 

We combine the branching fractions and the ratio of the $B^+$ and $B^0$ lifetime 
\hbox{$\tau_{+}/\tau_{0} = 1.071 \pm 0.009$~\cite{PDG:2008}} to obtain the 
isospin asymmetry \hbox{$\Delta_{0-} = \isoval \pm \isostat \pm \isosyst$}, 
which corresponds to \hbox{$\isolow < \Delta_{0-} < \isohi$} at the 90\% confidence interval.  We also measure $\acp(\bkpg) = \acpkp \pm \acpkpstat \pm \acpkpsys$.  The total combined $CP$ asymmetry is
$\acp = \acpval \pm \acpstat \pm \acpsyst$, with a 90\% confidence interval of $\acplolim < \acp < \acphilim$.

\begin{figure}[htpb]

\includegraphics[width=0.5\linewidth,clip=true]{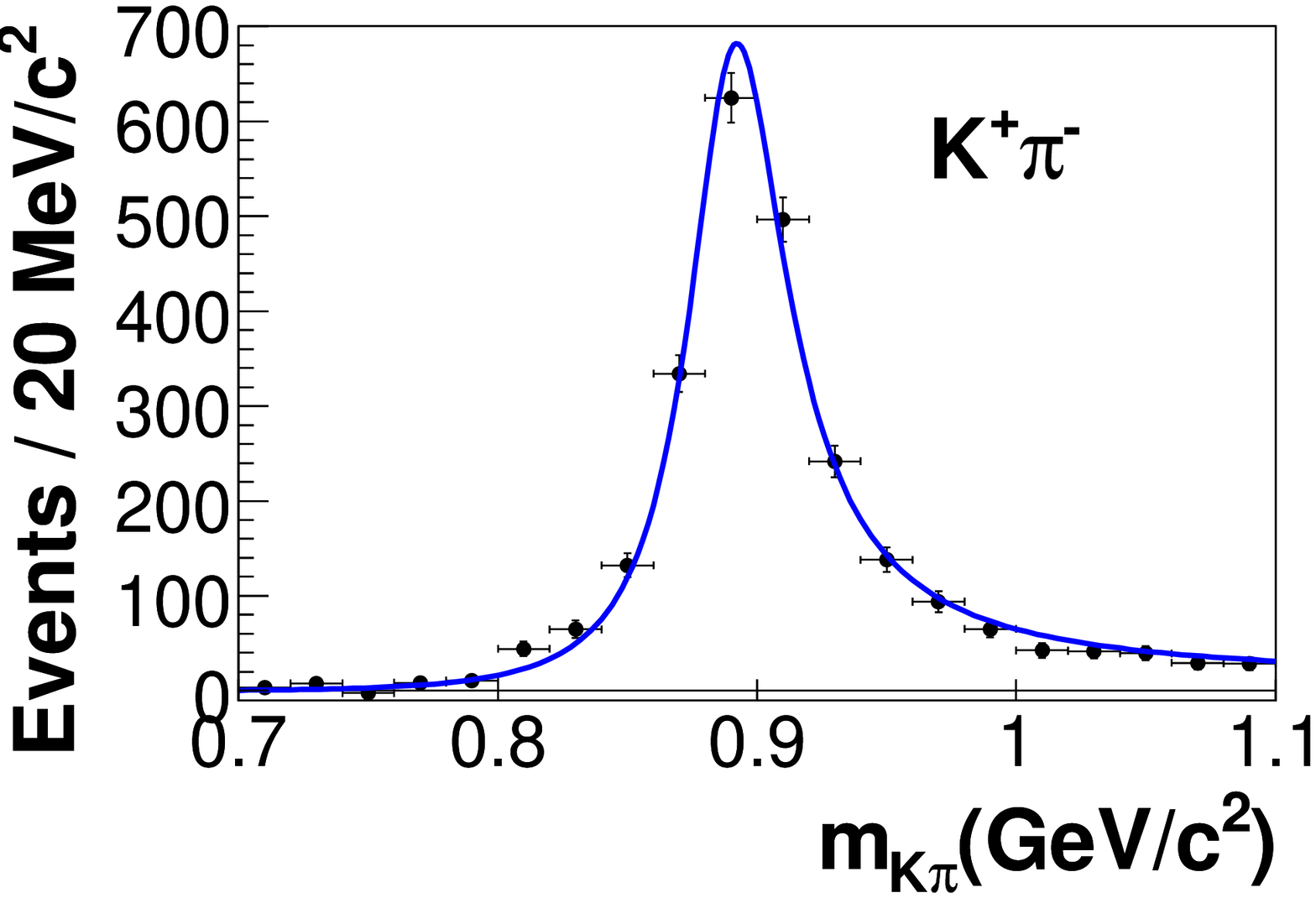}%
\includegraphics[width=0.5\linewidth,clip=true]{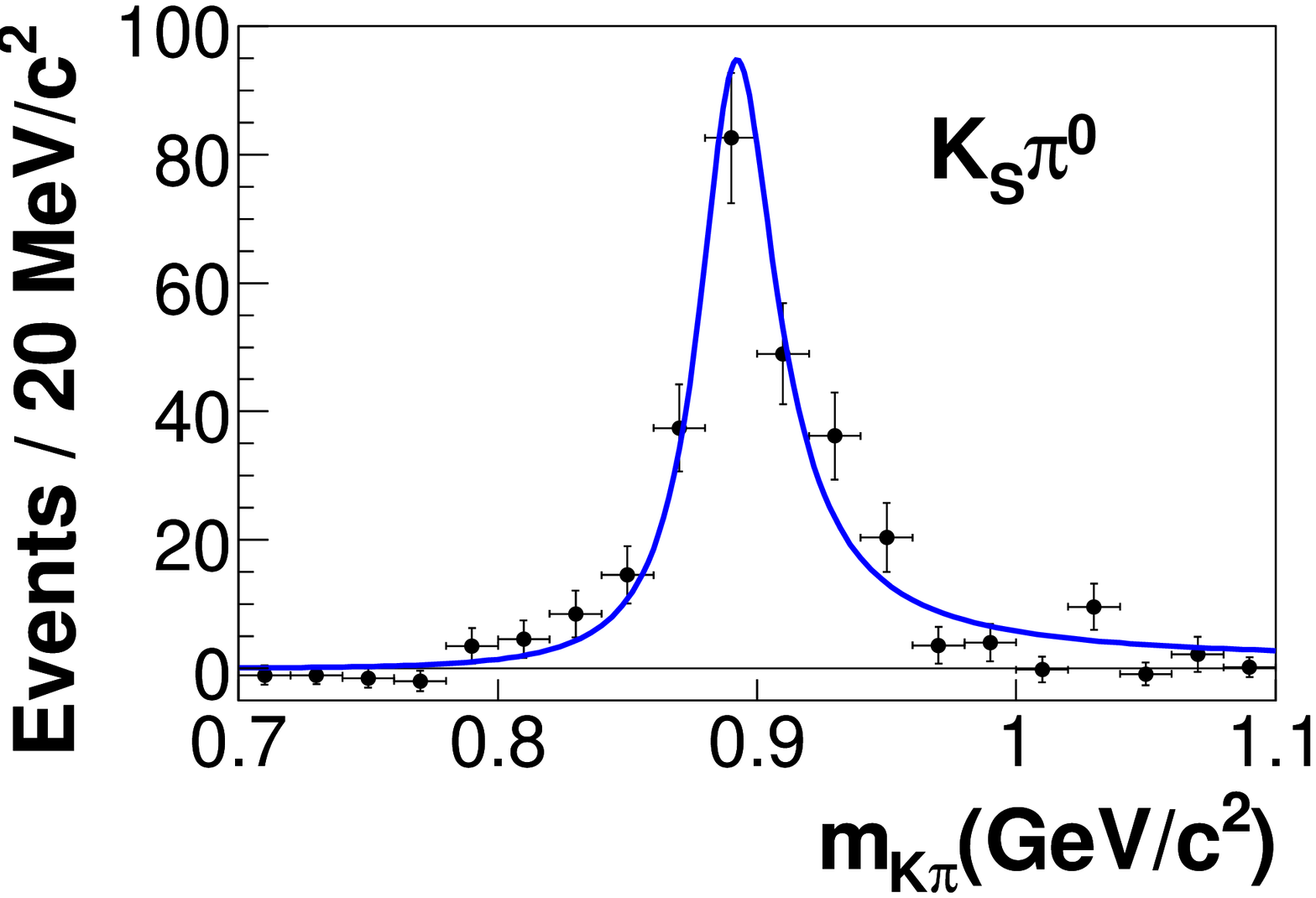}
\includegraphics[width=0.5\linewidth,clip=true]{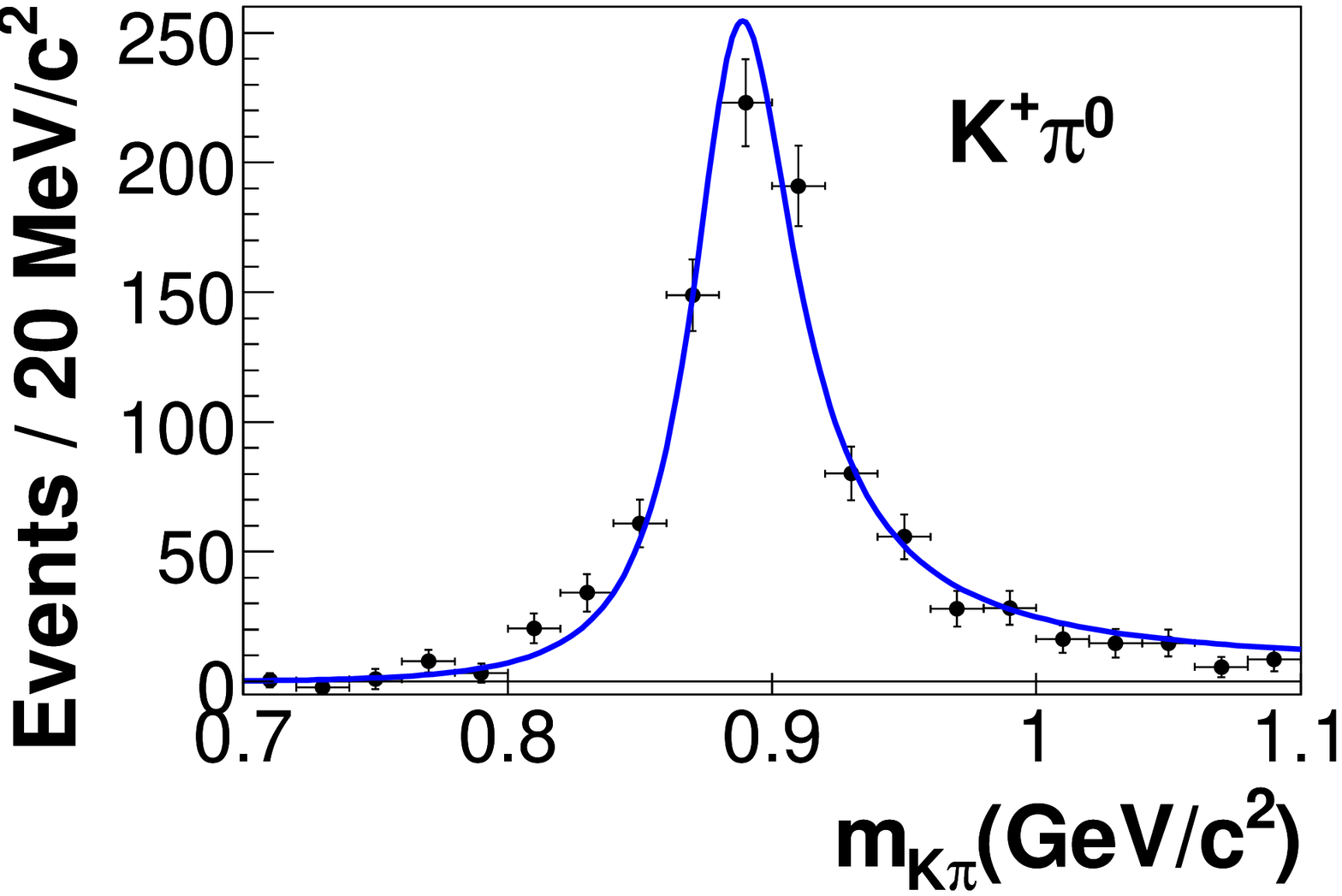}%
\includegraphics[width=0.5\linewidth,clip=true]{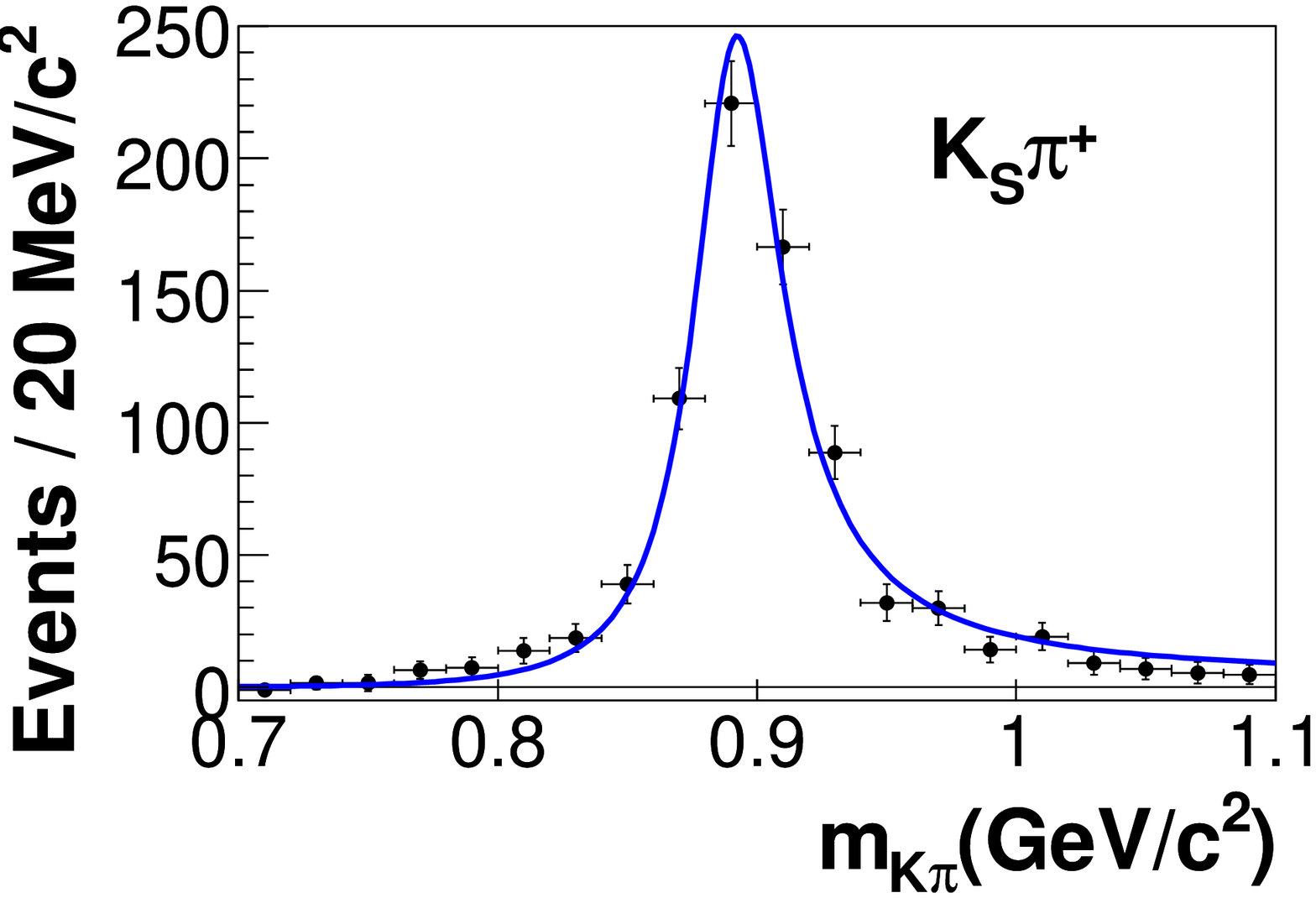}

\caption{ Fit of a single relativistic P-wave Breit-Wigner line shape (solid line) 
to the $K\pi$ invariant mass distribution of the sPlot of data (points).  For the $\kszksz$ and $\kspkpz$, the Breit-Wigner is convolved with a Gaussian of width 10 $\mev$.
}\label{fig:kstarmass}
\end{figure}

Figure~\ref{fig:kstarmass} shows the relativistic $P$-wave Breit-Wigner 
line shape fit to the sPlot~\cite{sPlot} of the $K\pi$ 
invariant mass distribution of data projecting out the signal component.  
For the $\kszksz$ and $\kspkpz$ modes, we convolve the Breit-Wigner 
line shape with a Gaussian with a width of 10 $\mev$ 
(determined from MC simulation) to account for detector resolution.  For the $\kszkp$ and the $\kspksp$ modes, the detector resolution is negligible.  The results are consistent with the signal events containing only P-wave $K^*$ mesons
and no other $K \pi$ resonances.

\begin{table}[htpb]

        \begin{center}

        \caption{Systematic errors (in \%) of the branching fractions.}
         \label{table:sys}
        \begin{tabular*}{\linewidth}{
@{\extracolsep{\fill}}l
@{\extracolsep{\fill}}c
@{\extracolsep{\fill}}c
@{\extracolsep{\fill}}c
@{\extracolsep{\fill}}c
}\hline\hline

Mode                            &    $K^+\pi^-$ & $K_S\pi^0$ & $K^+\pi^0$ & $K_S\pi^+$ \\\hline 
$\BR(\Upsilon(4S)\!\to\!\BzBzb)/\BR(\Upsilon(4S)\!\to\!\BpBm) $     &      1.6  &  1.6  &  1.6  &  1.6 \\
$B\overline{B}$ sample size     &      1.1  &  1.1  &  1.1  &  1.1 \\ 
Tracking efficiency             &      1.2  &  -    &  0.6  &  0.8 \\ 
Particle identification         &      0.6  &  -    &  0.6  &  0.2 \\ 
Photon selection                &      2.2  &  2.2  &  2.2  &  2.2 \\ 
$\pi^0$ reconstruction          &       -  &  3.0   &  3.0  &   -  \\
$\pi^0$ and $\eta$ veto         &      1.0  &  1.0  &  1.0  &  1.0 \\
$K_{S}$ reconstruction          &       -   &  0.7  &   -   &  0.7 \\
Neural Net efficiency           &      1.5  &  1.0  &  1.0  &  1.0 \\ 
Fit Model                       &      0.8  &  5.6  &  3.1  &  1.7 \\ 
Signal PDF bias                 &      0.9  &  2.2  &  1.6  &  1.4 \\\hline
Sum in quadrature               &      3.9  &  7.5  &  5.7  &  4.1  \\
\hline \hline
\end{tabular*}

\end{center}

\end{table}

We conclude that using a sample that is almost five times larger than previously used,
we have signficantly improved upon previous measurements of the $\bkg$ decay 
processes~\cite{Coan:1999kg,  Dasu:2004kg, Nakao:2004kg}.  
The measured isospin- and \CP-asymmetries and branching fractions are consistent with SM expectations.  

\input pubboard/acknow_PRL

\end{document}

%% file: pubboard/authors_mar2009.tex
%% author list as of 03-Mar-2009 (490 authors)
%
\author{B.~Aubert}
\author{Y.~Karyotakis}
\author{J.~P.~Lees}
\author{V.~Poireau}
\author{E.~Prencipe}
\author{X.~Prudent}
\author{V.~Tisserand}
\affiliation{Laboratoire d'Annecy-le-Vieux de Physique des Particules (LAPP), Universit\'e de Savoie, CNRS/IN2P3,  F-74941 Annecy-Le-Vieux, France}
\author{J.~Garra~Tico}
\author{E.~Grauges}
\affiliation{Universitat de Barcelona, Facultat de Fisica, Departament ECM, E-08028 Barcelona, Spain }
\author{M.~Martinelli$^{ab}$}
\author{A.~Palano$^{ab}$ }
\author{M.~Pappagallo$^{ab}$ }
\affiliation{INFN Sezione di Bari$^{a}$; Dipartimento di Fisica, Universit\`a di Bari$^{b}$, I-70126 Bari, Italy }
% ...

%% file: pubboard/acknow_PRL.tex
We are grateful for the excellent luminosity and machine conditions
provided by our \pep2\ colleagues, 
and for the substantial dedicated effort from
the computing organizations that support \babar.
The collaborating institutions wish to thank 
SLAC for its support and kind hospitality. 
This work is supported by
DOE
and NSF (USA),
NSERC (Canada),
CEA and
CNRS-IN2P3
(France),
BMBF and DFG
(Germany),
INFN (Italy),
FOM (The Netherlands),
NFR (Norway),
MES (Russia),
MEC (Spain), and
STFC (United Kingdom). 
Individuals have received support from the
Marie Curie EIF (European Union) and
the A.~P.~Sloan Foundation.

%% file: paper_la.bbl
\begin{thebibliography}{99}
\bibitem{CC} 
Charge conjugate modes are implied throughout, except for the \CP asymmetry.
\end{thebibliography}
